\documentclass[pdflatex,sn-mathphys-num]{sn-jnl}

\usepackage{graphicx}%
\usepackage{multirow}%
\usepackage{amsmath,amssymb,amsfonts}%
\usepackage{amssymb}
\usepackage{mathrsfs}
\usepackage{amsthm}%
\usepackage{mathrsfs}%
\usepackage[title]{appendix}%
\usepackage{xcolor}%
\usepackage{textcomp}%
\usepackage{manyfoot}%
\usepackage{booktabs}%
\usepackage{algorithm}%
\usepackage{algorithmicx}%
\usepackage{algpseudocode}%
\usepackage{listings}%
\usepackage{float}
\usepackage{placeins}
\usepackage{caption}
\usepackage{graphicx}
\usepackage{amsmath,amssymb}
\usepackage{booktabs}
\usepackage{multirow}

\usepackage{tikz}
\usepackage{pgfplots}
\pgfplotsset{compat=1.18}
\usetikzlibrary{shapes.geometric, arrows.meta, positioning, fit, calc}

\definecolor{eqblue}{RGB}{33, 64, 154}
\definecolor{eqred}{RGB}{150, 40, 40}
\definecolor{eqgray}{gray}{0.2}

\usepackage{caption}
\theoremstyle{thmstyleone}%
\theoremstyle{thmstyletwo}%

\theoremstyle{thmstylethree}%

\begin{document}

\title[Article Title]{Enhancing Adversarial Robustness in Network Intrusion Detection: A Layer-wise Adaptive Regularization Approach}


\author[1]{\fnm{Hira} \sur{Nasir}}\email{231289@students.au.edu.pk}

\author[1]{\fnm{Eiman} \sur{Javed}}\email{231293@students.au.edu.pk}

\author[1]{\fnm{Balawal} \sur{Shabir}}\email{balawal.shabir@au.edu.pk}

\author[1]{\fnm{Zunera} \sur{Jalil}}\email{zunera.jalil@au.edu.pk}

\author*[2]{\fnm{Ahmad} \sur{Mohsin}}\email{a.mohsin@ecu.edu.au}

\affil[1]{\orgdiv{National Cyber Security Academy (NCSA)}, \orgname{Air University}, \orgaddress{\city{Islamabad}, \country{Pakistan}}}

\affil*[2]{\orgdiv{School of Science}, \orgname{Edith Cowan University}, \orgaddress{\country{Australia}}}


\abstract{The new wave of adversarial attacks that utilize gradient-related vulnerabilities in neural network-based classifiers makes Network Intrusion Detection Systems more open to such threats. Although state-of-the-art adversarial training methods have shown promising results in producing more robust classifiers, their interpretability and defense ability are limited due to their lack of understanding of how adversarial attacks propagate in different layers of network classifiers. In this paper, we present an insightful approach, called LARAR (Layer-wise Adversarial Robustness using Adaptive Regularization), that incorporates additional layer-wise vulnerability analysis and adaptive weighting in conventional adversarial training methods. Additionally, we utilize 'Auxiliary Classifiers' in our approach. LARAR provides interpretable layer-wise vulnerability scores, achieves a clean accuracy of 95.01\%, and provides better robustness against adversarial attacks (FGSM, PGD, and transfer attacks) on the UNSW-NB15 dataset. Through the identification of vulnerable layers, the proposed framework reduces computational complexity and enables the early detection of adversarial samples, thus enhancing the effectiveness and interpretability of adversarial defense mechanisms in NIDS.}




\maketitle

\section{Introduction}\label{sec1}

{T}he level of increased advanced and sophisticated cyber threats has put Network Intrusion Detection Systems at the forefront of modern countermeasures against cybersecurity threats. The recent advancements in deep learning approaches have enabled Network Intrusion Detection Systems to achieve unparalleled accuracy in detecting network threats by automatically learning from the input data \cite{ennaji2025adversarial, hozouri2025comprehensive}. There has been an unforeseen weakness in incorporating machine learning models in applications related to security, due to an attack that adds perturbations to the original input while remaining imperceptible to human operators but causes the machine learning model to misclassify \cite{tafreshian2024defensive, kim2025enhancing}.

The consequences, if adversarial vulnerabilities in NIDS are exploited, are very critical. An adversary who can evade detection through adversarial perturbation can compromise network security while remaining undetected, which could lead to data breaches, disrupting services, or unauthorized access to critical infrastructures \cite{heydari2025enhancing}. Some recent works have shown that even state-of-the-art deep learning-based NIDS models are substantially vulnerable to white-box attacks such as the Fast Gradient Sign Method (FGSM) and Projected Gradient Descent (PGD), with attack success rates often exceeding 70\% \cite{awad2025enhanced, gurung2025enhancing, martinez2026adversarial}.

In response to such threats, a variety of defense strategies such as adversarial training, defensive distillation, and gradient masking have been proposed \cite{ndayipfukamiye2025adversarial, morshedi2025comprehensive}. Among all such adversarial training strategies, it has recently been shown that adversarial training is one of the most effective and theoretically justified defense strategies against adversarial attacks \cite{jamiri2025adversarial}. Recent studies have even proposed extending adversarial training with regularizers such as feature smoothing and gradient alignment, which impose constraints on models such that they are more resistant even to input perturbations \cite{chinnasamy2025deep,bhati2025neural}.

In spite of this, the existing methods in adversarial defense of NIDS have one important shortcoming: they assume that the neural network behaves like a ``black box,'' and their notion of robustness is based solely on top-layer prediction accuracy. Such an evaluation methodology does not give any information as to the flow of adversarial perturbations in the network's hidden layers, nor does it disclose the particular layers that are more vulnerable to this type of attack, as well as the layers where this might trigger a destabilization of the network's internal representation \cite{singh2025shallowest}. They apply a uniform regularization across all layers without any consideration that, in fact, the degree of vulnerability of a layer can dictate the effect of adversarial changes applied to that layer.

This lack of layer-wise analysis represents both a scientific gap in understanding adversarial robustness and a practical limitation in developing efficient defense mechanisms. Scientifically, only by understanding how adversarial perturbations accumulate and transform while propagating through neural network layers is it possible to develop principled defenses \cite{bereska2024mechanistic}. From a practical perspective, not having layer-specific vulnerability analysis prevents the conduction of targeted defense strategies, which would deploy most of the computational resources on the most vulnerable parts of the network.

This paper addresses these limitations by introducing \textbf{LARAR}, a framework that extends adversarial training with three key innovations:

\begin{enumerate}
  \item \textbf{Layer Vulnerability Score (LVS) Metric:}
  We propose to define a new quantified metric that measures the sensitivity of each hidden layer to adversarial attacks by computing the normalized difference between clean and adversarial activations. This enables the identification of vulnerable layers within a given model architecture.

  \item \textbf{Adaptive Layer-wise Regularization:}
  Unlike existing approaches, \textbf{LARAR} employs a regularization mechanism with learnable weights, allowing the model to adaptively adjust the level of adversarial regularization according to the estimated vulnerability level. This effectively channels computational resources to the most critical and vulnerable parts of the network.

  \item \textbf{Multi-level Supervision with Early Detection:}
  To further enhance robustness, we introduce additional classifiers at intermediate layers of the network. This multi-level supervision enables earlier detection of adversarial behavior before the final classification layer is reached.
\end{enumerate}

The remainder of this paper is organized as follows: Section \ref{rw} reviews related work on adversarial robustness in NIDS and layer-wise analysis techniques. Section \ref{bp} provides background on adversarial attacks and the baseline ADVNN architecture. Section \ref{pfm} focuses on the problem and the limitations. Section \ref{plf} presents the \textbf{LARAR} framework. Section \ref{ese} describes our experimental setup and evaluation methodology. Section \ref{rad} presents comprehensive experimental results and analysis. Section \ref{c} concludes the paper.

\section{Related Work}\label{rw}
This section reviews the existing literature on adversarial attacks and defense mechanisms, with particular focus on their application to network intrusion detection systems.

\subsection{Adversarial Attacks on Machine Learning Systems}\label{subsec6}
Adversarial attacks on machine learning models have been extensively studied since their discovery by Szegedy et al. \cite{szegedy2013intriguing}, who demonstrated that imperceptible perturbations could cause misclassification in deep neural networks. Goodfellow et al. \cite{goodfellow2014explaining} introduced the FGSM, a computationally efficient single-step attack that exploits the gradient of the loss function. Building upon this work, Madry et al. \cite{madry2017towards} proposed PGD, an iterative variant that represents one of the strongest first-order adversarial attacks.

In the domain of network intrusion detection, adversarial attacks pose unique challenges compared to computer vision applications. Rigaki and Garcia \cite{rigaki2018bringing} demonstrated that gradient-based attacks could evade deep learning-based intrusion detection systems with minimal perturbations to network traffic features. Yang et al. \cite{yang2018adversarial} showed that adversarial examples generated for one NIDS model could transfer to other architectures.

\subsection{Adversarial Defense and Robustness in Network Intrusion Detection}
Defense strategies against adversarial attacks include adversarial training, defensive distillation, and ensemble-based defenses. Papernot et al. \cite{papernot2016distillation} proposed defensive distillation to reduce model sensitivity to input perturbations. Tram\`er et al. \cite{tramer2017ensemble} introduced ensemble adversarial training to improve robustness, while Zhang et al. \cite{zhang2019theoretically} analyzed the fundamental trade-off between robustness and accuracy.\\

A thorough analysis of adversarial machine learning in NID was carried 
out by Alhajjar et al. \cite{alhajjar2021adversarial} and Han et al.
\cite{han2021evaluating} who proposed a domain-specific adversarial 
training framework for network traffic classification but did not 
investigate layer-wise vulnerabilities. Zhang et al. 
\cite{zhang2022adversarial} further demonstrated that gradient-based 
attacks can be systematically crafted against deep learning-based NIDS 
and proposed several defense mechanisms, highlighting the need for 
principled robustness evaluation.
\subsection{Layer-Wise Analysis in Deep Learning}
Raghu et al. \cite{raghu2017svcca} introduced SVCCA to analyze layer representations in deep neural networks. Ilyas et al. \cite{ilyas2019adversarial} attributed adversarial vulnerability to non-robust features learned by networks but did not provide layer-specific vulnerability quantification. These studies highlight that adversarial vulnerability is intrinsically linked to internal feature representations. While these approaches provide valuable insights into representation learning and feature vulnerability, they do not offer a systematic framework for layer-wise robustness assessment and mitigation during training.

\subsection{Comparative Analysis}\
Table \ref{tab:related_work_comparison} provides a detailed comparison between various existing adversarial defense approaches for NIDS with their respective strengths and limitations, along with key differentiators of our proposed framework, \textbf{LARAR.} Unlike prior works, which employ either uniform defense mechanisms or focus only on the output level, our approach presents layer-wise vulnerability quantification and adaptive regularization of adversarial defense, thus providing interpretability with efficiency specific to network intrusion detection.\\

From Table \ref{tab:related_work_comparison}, it is clear that the current adversarial defense mechanisms for NIDS have three major limitations: (1) no layer-wise vulnerability analysis that can provide targeted defense strategies; (2) no adaptive regularization mechanism that dynamically gives more importance to the vulnerable components; and (3) lack of interpretability about internal adversarial behavior. Our proposed \textbf{LARAR} framework directly fills these gaps by introducing Layer Vulnerability Scores for quantifying internal robustness, learnable adaptive weights for layer-specific regularization, and auxiliary classifiers for multi-level supervision. The holistic approach not only boosts adversarial robustness but also presents actionable insights for security practitioners to comprehend and mitigate adversarial threats at a granular level.
\FloatBarrier
\begin{table*}[!htbp]
\centering
\small
\renewcommand{\arraystretch}{2.0}
\setlength{\tabcolsep}{3pt}

\resizebox{\textwidth}{!}{%
\begin{tabular}{|p{3.0cm}|c|p{3.5cm}|c|c|p{2.0cm}|c|p{3.5cm}|}
\hline
\textbf{Approach} & \textbf{Year} & \textbf{Defense Strategy} & \textbf{Layer-Wise} & \textbf{Adaptive Reg.} & \textbf{Dataset} & \textbf{Interpretability} & \textbf{Key Limitation} \\ 
\hline\hline
\multicolumn{8}{|c|}{\textbf{Early Adversarial Training Methods}} \\
\hline
Goodfellow et al.\ \cite{goodfellow2014explaining} & 2015 & Standard Adversarial Training & No & No & MNIST/ CIFAR & Low & Uniform defense, no layer insights \\ 
\hline
Madry et al.\ \cite{madry2017towards} & 2018 & PGD-based Adversarial Training & No & No & CIFAR-10 & Low & Black-box approach, high compute cost \\ 
\hline
Papernot et al.\ \cite{papernot2016distillation} & 2016 & Defensive Distillation & No & No & MNIST & Low & Vulnerable to adaptive attacks \\ 
\hline
Tram\`er et al.\ \cite{tramer2017ensemble} & 2018 & Ensemble Adversarial Training & No & No & CIFAR-10 & Medium & High resource needs, no layer analysis \\ 
\hline\hline
\multicolumn{8}{|c|}{\textbf{NIDS-Specific Adversarial Defense (2021--2023)}} \\
\hline
Rigaki \& Garcia \cite{rigaki2018bringing} & 2018 & Input Validation & No & No & CTU-13 & Low & Limited to known attack patterns \\ 
\hline
Han et al.\ \cite{han2021evaluating} & 2021 & Domain-Specific Adversarial Training & No & No & NSL-KDD & Low & Treats model as black box \\ 
\hline
Alhajjar et al.\ \cite{alhajjar2021adversarial} & 2021 & Survey and Taxonomy & N/A & N/A & N/A & N/A & No defense mechanism proposed \\ 
\hline
Yuan et al.\ \cite{yuan2024simple} & 2023 & Hybrid DL+ML with AE detector & No & No & NSL-KDD & Medium & No layer vulnerability quantification \\ 
\hline
Xiong et al.\ \cite{xiong2023aidtf} & 2023 & GAN-based Adversarial Framework & No & No & CICIDS2017 & Medium & Computationally expensive, no interpretability \\ 
\hline\hline
\multicolumn{8}{|c|}{\textbf{Recent State-of-the-Art NIDS Defenses (2024--2025)}}\\
\hline
Heydari et al.\ \cite{heydari2025enhancing} & 2025 & Composite Adversarial Training (ADV-NN) & No & No & UNSW-NB15 & Low & Uniform regularization across all layers \\ 
\hline
Patel et al.\ \cite{gurung2025enhancing} & 2025 & FGSM Training on XGBoost & No & No & NF-ToN-IoT & Low & ML-based, no deep layer analysis \\ 
\hline
Ahmad et al.\ \cite{awad2025enhanced} & 2025 & Ensemble Defense Framework & No & No & CICIDS 2017-18 & Medium & Ensemble complexity, no layer insights \\ 
\hline
Ilyas et al.\ \cite{ilyas2019adversarial} & 2019 & Non-robust Feature Analysis & Partial & No & ImageNet & High & No layer-specific vulnerability quantification \\ 
\hline\hline
\textbf{LARAR (Proposed)} & \textbf{2026} & \textbf{Layer-Wise Adaptive Training} & \textbf{Yes} & \textbf{Yes} & \textbf{UNSW-NB15} & \textbf{High} & \textbf{Requires layer-wise computation (+18\% overhead)} \\ 
\hline
\end{tabular}%
}
\caption{Comparative Analysis of Adversarial Defense Approaches for Network Intrusion Detection Systems}
\label{tab:related_work_comparison}
\end{table*}
\FloatBarrier
\vspace*{1mm}

\subsection{Research Gap}
Current adversarial defense methods for NIDS have three key limitations that are addressed by our solution. Firstly, existing methods uniformly regulate all layers over the network without taking into consideration that various layers have completely different susceptibility patterns to adversarial attacks based \cite{madry2017towards}. Secondly, existing methods lack insight within the network by regarding models as a black box without studying the perturbation propagation and amplification processes over various layers \cite{alhajjar2021adversarial}. This makes it difficult for them to identify which layers of the network are more susceptible to attacks and therefore require more resources for defense\cite{han2021evaluating}. Thirdly, existing methods overlook the early adversarial detection process based on information over various layers to gain efficiency through early exit strategies based on computational implementations \cite{raghu2017svcca,chen2023layer}.\\

The gaps in understanding and analysis of adversarial robustness in neural networks are filled with the introduction of three innovations in our proposed framework, \textbf{LARAR}. We incorporate LVS, a metric to quantify and analyze adversarial influenced layers of a neural net on an interpretable scale. We also incorporate an adaptive regulation technique with weights to learn and give importance to adversarial vulnerable layers. In our work, we add additional classifiers for multi-level training, further demonstrating the possibility of an earlier detection system for adversarial examples in intermediate layers. In contrast to previous works that have concentrated primarily on output levels for NIDS robustness, our proposal, \textbf{LARAR}, is a major improvement in adversarial NIDS, as it offers interpretable results and scope for efficiencies.

\section{Background and Preliminaries}\label{bp}
This section establishes the formal mathematical framework for network intrusion detection and adversarial attacks.

\subsection{Network Intrusion Detection Setting}
We consider a binary classification problem for network intrusion detection where samples from the feature space $\mathcal{X} \subseteq \mathbb{R}^d$ are mapped to labels $\mathcal{Y} = \{0, 1\}$, where 0 represents normal traffic and 1 represents malicious activity. The NIDS model is a deep neural network $f_\theta: \mathcal{X} \rightarrow [0,1]$ parameterized by $\theta$, trained on a dataset $\mathcal{D} = \{(x_i, y_i)\}_{i=1}^N$ where $x_i \in \mathcal{X}$ represents network flow features and $y_i \in \mathcal{Y}$ represents the corresponding label.

The network architecture consists of $L$ hidden layers with activation functions, formally defined as:
\begin{align}
h^{(0)} &= x \\
h^{(l)} &= \sigma(W^{(l)} h^{(l-1)} + b^{(l)}), \quad l = 1, \ldots, L \\
\hat{y} &= f_\theta(x) = \text{sigmoid}(W^{(out)} h^{(L)} + b^{(out)})
\end{align}
where $h^{(l)}$ denotes the activation at layer $l$, $\sigma$ is the ReLU activation function, and $W^{(l)}, b^{(l)}$ are the weights and biases at layer $l$. In our implementation, we use $L=2$ hidden layers with dimensions $d_1=128$ and $d_2=64$.

The UNSW-NB15 dataset used in our experiments contains $d=42$ features after preprocessing, including flow-based statistics, packet-level characteristics, and protocol information \cite{moustafa2015unsw}. Features are normalized using standard scaling to ensure zero mean and unit variance, a critical preprocessing step for stable neural network training.

\subsection{Adversarial Attack Model}\
We consider an adversary with white-box access to the model parameters $\theta$, representing the strongest possible threat model \cite{yang2018adversarial}. The adversary aims to craft adversarial examples $x^{adv}$ that cause misclassifications while remaining within a bounded perturbation radius $\epsilon$. The adversarial perturbation $\delta$ is constrained by an $\ell_\infty$ norm:
\begin{equation}
x^{adv} = x + \delta, \quad \|\delta\|_\infty \leq \epsilon
\end{equation}

The $\ell_\infty$ constraint ensures that no single feature is perturbed by more than $\epsilon$, maintaining the semantic validity of network traffic in the NIDS context \cite{zhang2022adversarial}. We focus on two primary attack strategies commonly used in adversarial robustness evaluation:

\textbf{Fast Gradient Sign Method (FGSM):} \cite{goodfellow2014explaining} A single-step attack that perturbs the input in the direction of the gradient:
\begin{equation}
x^{adv} = x + \epsilon \cdot \text{sign}(\nabla_x \mathcal{L}(f_\theta(x), y))
\end{equation}
where $\mathcal{L}$ is the binary cross-entropy loss. FGSM is computationally efficient and serves as a baseline for measuring basic adversarial robustness.

\textbf{Projected Gradient Descent (PGD):} \cite{madry2017towards} An Iterative Method
that applies multiple gradient steps with projection onto the $\epsilon$-ball:
\begin{align}
x^{adv}_0 &= x + \text{Uniform}(-\epsilon, \epsilon) \\
x^{adv}_{t+1} &= \Pi_{\mathcal{B}_\epsilon(x)}\left(x^{adv}_t + \alpha \cdot \text{sign}(\nabla_x \mathcal{L}(f_\theta(x^{adv}_t), y))\right)
\end{align}
where $\Pi_{\mathcal{B}_\epsilon(x)}$ projects onto the $\ell_\infty$ ball of radius $\epsilon$ at $x$, where $\alpha$ is the step size (usually $\alpha = \epsilon/K$ where $K$ is the number of iterations), and random initialization within the $\epsilon$-ball ensures the attack explores the perturbation space effectively. PGD signify a strong first-order adversary and is thought to be one of the most powerful techniques for evaluating adversarial robustness.

We also assess the transfer attacks, which are the adversarial examples produced on the surrogate model are tested on our
target model, simulating black-box attack scenarios where the
adversary does not have direct access to model parameters.

\section{Problem Formulation and Motivation}\label{pfm}
This section introduces the LVS metric and formulates the extended optimization objective that enables adaptive layer-wise defense.

\subsection{Theoretical Basis for LVS}
The LVS metric is grounded in the theory of adversarial sensitivity in deep networks. Formally, a layer $l$ is considered vulnerable if small $\ell_\infty$-bounded input perturbations $\delta$ produce
disproportionately large shifts in its activations. Prior work has shown that adversarial perturbations tend to amplify as they propagate through successive nonlinear layers, particularly in layers lacking strong implicit regularization \cite{madry2017towards, zhang2019theoretically}. Layer-wise analysis of these amplification dynamics has been studied through the lens of singular value decomposition of weight matrices, where layers with larger spectral norms exhibit higher sensitivity to input perturbations \cite{bartlett2017spectrally}. Our LVS metric operationalizes this insight as an empirical, batch-averaged sensitivity measure that is architecture-agnostic and computationally lightweight.

\subsubsection{Perturbation Propagation Across Depth}

Let $\delta^{(l)} = h^{(l)}(x^{adv}) - h^{(l)}(x)$ denote the activation shift at layer $l$ induced by an adversarial perturbation $\delta = x^{adv} - x$. For a ReLU network, the propagated shift satisfies:
\begin{equation}
    \|\delta^{(l)}\|_2 \leq \sigma_{\max}(W^{(l)}) \cdot
    \|\delta^{(l-1)}\|_2
    \label{eq:propagation}
\end{equation}
where $\sigma_{\max}(W^{(l)})$ denotes the largest singular value of the weight matrix $W^{(l)}$ at layer $l$, also known as its spectral norm. Intuitively, $\sigma_{\max}(W^{(l)})$ measures the maximum factor by which the linear transformation $W^{(l)}$ can stretch a vector. When $\sigma_{\max}(W^{(l)}) > 1$, the layer \textit{amplifies} the incoming perturbation $\|\delta^{(l-1)}\|_2$; when $\sigma_{\max}(W^{(l)}) < 1$, it \textit{attenuates} it.
This recurrence reveals that adversarial perturbations do not propagate uniformly rather they grow or shrink layer-by-layer depending on each layer's spectral norm, which directly motivates layer-specific rather than uniform regularization. The LVS, defined formally in Section~\ref{sec:lvs_def}, provides a batch-averaged
empirical estimate of $\|\delta^{(l)}\|_2 / \|h^{(l)}(x)\|_2$, serving as a practical and architecture-agnostic surrogate.

\subsection{Layer Vulnerability Analysis}\label{sec:lvs_def}
To measure the effect of adversarial perturbations on individual internal representations, we define the Layer Vulnerability Score (LVS):

\begin{equation}
\text{LVS}^{(l)} = \frac{1}{B} \sum_{i=1}^B \frac{\|h^{(l)}(x_i^{adv}) - h^{(l)}(x_i)\|_2}{\|h^{(l)}(x_i)\|_2 + \epsilon_{small}}
\label{eq:lvs}
\end{equation}

where $h^{(l)}(x)$ represents the activation at layer $l$ for input $x$, $B$ is the batch size, and $\epsilon_{small} = 10^{-8}$ prevents division by zero. The LVS measures the relative change in layer activations caused by adversarial perturbations, normalized by the amount of clean activations to ensure invariance with respect to scales across layers with different dimensionalities.

When LVS value is high it indicates that layer $l$ significantly amplifies adversarial perturbations, making it a critical vulnerability point. On the contrary, if the level of security vulnerability is low, this indicates that the layer is robust naturally, perhaps because of implicit regularization during training.

The LVS metric also lends itself to several practical applications
\begin{itemize}
\item \textbf{Vulnerability identification:} Pinpoints the most manipulative levels and guides architects on how to improve the design.
\item \textbf{Perturbation propagation analysis:} Reveals how adversarial noises are accumulated or dissipated with increasing depth of the network.
\item \textbf{Defense prioritization:} Enables targeted allocation of regularization resources to vulnerable layers.
\item \textbf{Early detection capability:} Supports mechanisms for thresholding adversarial detection at intermediate levels.
\end{itemize}

Unlike global robustness metrics, which only account for the final output perturbations, LVS has fine-grained internal visibility that is highly essential in understanding and improving adversarial defenses.

\subsection{Problem Formulation}\label{subsec8}
Traditional adversarial training solves the following min-max optimization problem:
\begin{equation}
\min_\theta \mathbb{E}_{(x,y) \sim \mathcal{D}} \left[ \max_{\|\delta\|_\infty \leq \epsilon} \mathcal{L}(f_\theta(x + \delta), y) \right]
\label{eq:traditional}
\end{equation}

This formulation learns the model to minimize worst-case loss under adversarial perturbation. However, it treats all the layers uniformly without providing a mechanism for layer-specific defense adaptation. The inner maximization finds the strongest attack within the perturbation budget, whereas the outer minimization adjusts the model parameters to be robust against such an attack.

We extend this formulation to include layer-wise vulnerability analysis and adaptive regularization:

\begin{align}
\min_\theta \min_{w^{(1)}, \ldots, w^{(L)}} \mathbb{E}_{(x,y) \sim \mathcal{D}} \Bigg[ &\max_{\|\delta\|_\infty \leq \epsilon} \mathcal{L}_{total}(x, x+\delta, y; \theta, w) \Bigg]
\end{align}

where the total loss function is:
\begin{align}
\mathcal{L}_{total} &= \mathcal{L}_{CE} + \lambda_{aux}\mathcal{L}_{aux} + \lambda_{GA} \mathcal{L}_{GA} \nonumber \\
&\quad + \lambda_{FS} \mathcal{L}_{FS} + \beta \sum_{l=1}^L w^{(l)} \cdot \text{LVS}^{(l)}
\label{eq:proposed}
\end{align}

The components of this composite loss are designed to address different aspects of adversarial robustness:

\textbf{Cross-Entropy Loss ($\mathcal{L}_{CE}$):} Standard classification loss computed on both clean and adversarial examples:
\begin{equation}
\mathcal{L}_{CE} = \frac{1}{2}\left[\mathcal{L}_{BCE}(f_\theta(x), y) + \mathcal{L}_{BCE}(f_\theta(x^{adv}), y)\right]
\end{equation}
This ensures the model maintains accuracy on clean data while learning to classify adversarial examples correctly.

\textbf{Auxiliary Loss ($\mathcal{L}_{aux}$):} Multi-level supervision through auxiliary classifiers attached to each hidden layer:
\begin{equation}
\mathcal{L}_{aux} = \sum_{l=1}^L \mathcal{L}_{BCE}(g^{(l)}(h^{(l)}(x)), y)
\end{equation}
where $g^{(l)}: \mathbb{R}^{d_l} \rightarrow [0,1]$ is an auxiliary classifier (single linear layer) attached to layer $l$. This strengthens intermediate representations by providing direct supervision at each layer, reducing internal drift caused by adversarial perturbations.

\textbf{Gradient Alignment Loss ($\mathcal{L}_{GA}$):} Ensures consistent gradient geometry between clean and adversarial samples:
\begin{equation}
\mathcal{L}_{GA} = \mathbb{E}\left[\left\|\nabla_x \mathcal{L}_{BCE}(f_\theta(x), y) - \nabla_x \mathcal{L}_{BCE}(f_\theta(x^{adv}), y)\right\|_2^2\right]
\end{equation}
Adversarial attacks use the strong gradients to find effective attacks. In our approach, we align the gradients of the adversarial and clean examples. It makes the adversarial attack harder because an effective attack requires smoothed gradients.

\textbf{Feature Smoothing Loss ($\mathcal{L}_{FS}$):} Encourages Stability in Deep Layer Representations:
\begin{equation}
\mathcal{L}_{FS} = \|h^{(L)}(x) - h^{(L)}(x^{adv})\|_2^2
\end{equation}
This regularizes the last hidden layer so it generates similar output representations for both clean and adversarial inputs, which reduces the model's sensitivity to changes in the feature space \cite{papernot2016distillation}.

\textbf{Layer-Wise LVS Regularization:} Adaptive penalty based on measured layer vulnerability:
\begin{equation}
\mathcal{L}_{LVS} = \beta \sum_{l=1}^L w^{(l)} \cdot \text{LVS}^{(l)}
\end{equation}
where $w^{(l)} \in \mathbb{R}^+$ are learnable layer-specific weights initialized to 1.0. Unlike fixed regularization, these weights adapt during training via gradient descent:
\begin{equation}\label{is}
w^{(l)} \leftarrow w^{(l)} - \eta \frac{\partial \mathcal{L}_{total}}{\partial w^{(l)}}
\end{equation}
This allows the optimization process to automatically learn over which layers to perform more intense regularization, since empirical vulnerability will serve as a pointer to that, and therefore distribute defensive resources more effectively.

The hyperparameters $\lambda_{aux}=0.2$, $\lambda_{GA}=1.0$, $\lambda_{FS}=0.5$, and $\beta=0.3$ control the relative importance of each loss component and were determined through validation experiments.

The objective in Equation \ref{eq:proposed} gives a novel formulation that links adversarial training, interpretable layer-wise analysis, and adaptive regularization. Unlike traditional methods that uniformly apply defensive strategy,\textbf{ LARAR }allows the model to allocate defensive resources according to the empirically measured vulnerability in each layer.

\section{Proposed LARAR Framework}\label{plf}
This section presents the complete \textbf{LARAR} architecture and it's key components.

\subsection{Architecture Overview}\label{subsec10}
\textbf{LARAR} is an extension to the general model of neural networks with the following enhancements: layer vulnerability monitoring, adaptive regularization, as well as an early detection module to be incorporated into the general feed-forward model of the networks.

The base network is composed of two dense hidden layers of dimensions 128 and 64. In addition, ReLU nonlinearities and batch normalization techniques were employed for stability. The two dense hidden layers were further enhanced with the following components:

\begin{itemize}
\item \textbf{Activation hooks} that capture layer outputs for both clean and adversarial inputs during the forward pass, allowing for real-time vulnerability assessment.
\item \textbf{Auxiliary classifiers} that are single linear layers providing intermediate supervision signals which strengthen feature representations.
\item \textbf{Vulnerability monitors} that compute LVS values through clean versus adversarial activations at each layer.
\item \textbf{Detection thresholds} ($\tau^{(l)}$) that flag the potential inputs that are adversarial if LVS outperform pre-defined limits.
\end{itemize}
\FloatBarrier
\begin{figure}[!htbp]
\centering
\includegraphics[width=\columnwidth]{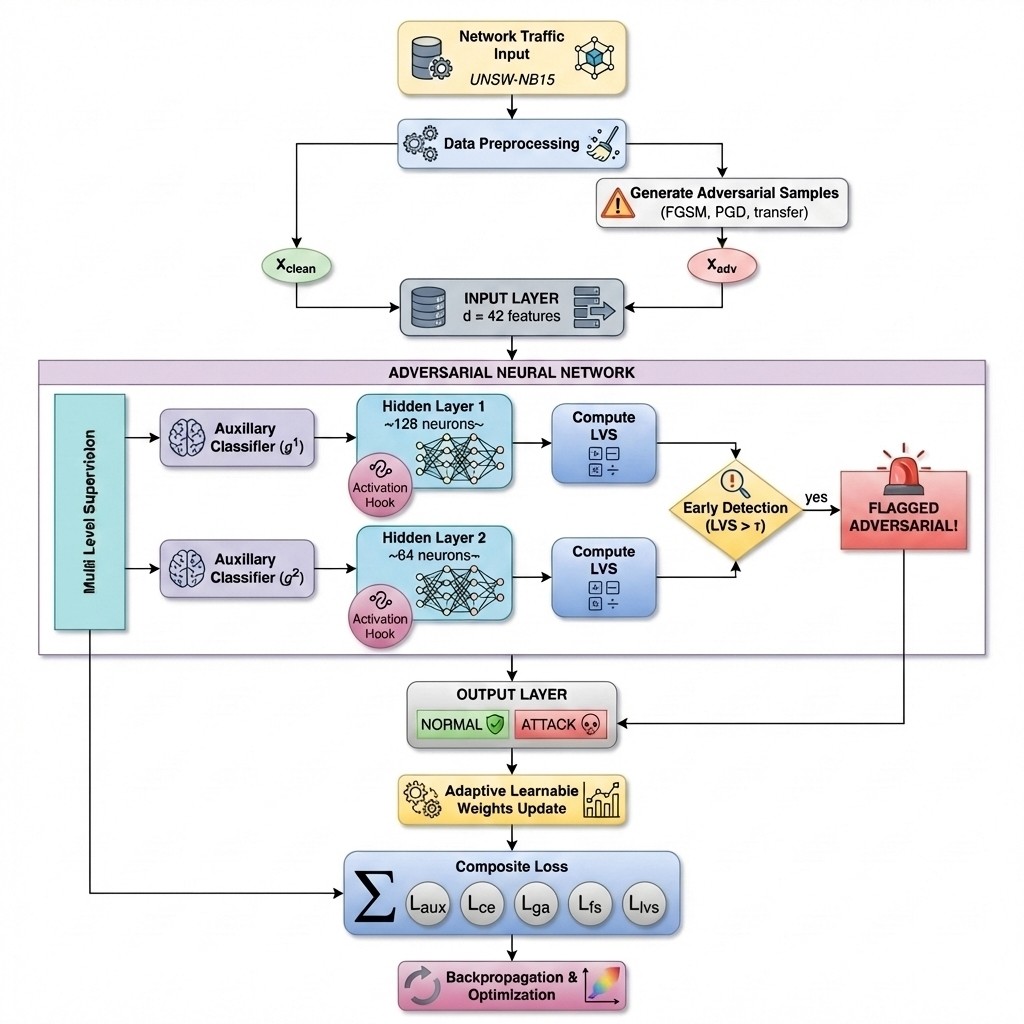}
\caption{LARAR architecture.}
\label{fig:larar_architecture}
\end{figure}
\FloatBarrier
\vspace*{5mm}
The forward pass processes inputs through the network while simultaneously recording activations at each layer. During training, both clean samples $x$ and their adversarial counterparts $x^{adv}$ (generated via PGD attack) are forwarded through the network, allowing the computation of layer-wise vulnerability scores directly. Figure~\ref{fig:larar_architecture} illustrates the complete end-to-end \textbf{LARAR} pipeline.

\subsection{Layer-wise Vulnerability Quantification}
Algorithm \ref{alg:lvs} describes the detailed computation of Layer Vulnerability Scores during the training process.
\FloatBarrier
\begin{algorithm}[!htbp]
\caption{Layer Vulnerability Score Calculator}
\label{alg:lvs}
\begin{algorithmic}[1]

\State \textbf{Input:} Clean batch $X$, adversarial batch $X^{adv}$, model $f_\theta$, validation set $\mathcal{V}_{clean}$
\State \textbf{Output:} Layer vulnerability scores $\{\text{LVS}^{(l)}\}_{l=1}^L$

\State Initialize: $\mathcal{H}_{clean} = \{\}, \mathcal{H}_{adv} = \{\}$

\State Forward pass: $\hat{y}_{clean} = f_\theta(X)$
\For{each layer $l = 1$ to $L$}
    \State Store: $\mathcal{H}_{clean}^{(l)} = h^{(l)}(X)$
\EndFor

\State Forward pass: $\hat{y}_{adv} = f_\theta(X^{adv})$
\For{each layer $l = 1$ to $L$}
    \State Store: $\mathcal{H}_{adv}^{(l)} = h^{(l)}(X^{adv})$
\EndFor

\For{each layer $l = 1$ to $L$}
    \State $\Delta^{(l)} = \mathcal{H}_{adv}^{(l)} - \mathcal{H}_{clean}^{(l)}$
    
    \State $\text{LVS}^{(l)} = \frac{1}{B}\sum_{i=1}^B \frac{\|\Delta^{(l)}_i\|_2}{\|\mathcal{H}_{clean,i}^{(l)}\|_2 + 10^{-8}}$
    
    \State $\tau^{(l)} = \max\left\{ \mu_{clean}^{(l)} + k\sigma_{clean}^{(l)},\ \lambda \cdot \text{LVS}^{(l)}_{max} \right\}$
    
    \If{$\text{LVS}^{(l)} > \tau^{(l)}$}
        \State Flag batch as potentially adversarial
    \EndIf
\EndFor

\State \textbf{return} $\{\text{LVS}^{(l)}\}_{l=1}^L$

\end{algorithmic}
\end{algorithm}
\FloatBarrier
\vspace*{1mm}

\subsection{Adaptive Layer-Wise Regularization}
Traditionally, robust adversarial training employs a uniform regularization across layers, whereby each component is treated equally despite differences in its true vulnerability. \textbf{LARAR} employs learnable weights per layer $\{w^{(l)}\}_{l=1}^L$ to automatically focus on vulnerable layers throughout the course of the training process

The layer-wise regularization term is:
\begin{equation}\label{so}
\mathcal{L}_{LARAR} = \beta \sum_{l=1}^L w^{(l)} \cdot \text{LVS}^{(l)}
\end{equation}

where $\beta=0.3$ is a global scaling factor that controls the overall strength of layer-wise regularization relative to other loss components. The weights $w^{(l)}$ are implemented as learnable parameters, initialized to 1.0, and updated via gradient descent alongside model parameters $\theta$.

The gradient of the total loss with respect to layer weights is:
\begin{equation}
\frac{\partial \mathcal{L}_{total}}{\partial w^{(l)}} = \beta \cdot \text{LVS}^{(l)}
\end{equation}

If a particular layer $l$ has a high vulnerability throughout the back-propagation process, i.e., (high $\text{LVS}^{(l)}$),  the gradients would be driven further up the hill, thus enhancing the regularization penalty. On the contrary, naturally resilient layers with a high value of LVS would mean weakening the regularization weight over time, therefore leaving more regularization capacity for the more vulnerable layers.

\subsection{Multi-Level Supervision via Auxiliary Classifiers}
To enrich intermediate representational content and add supplementary gradient information to earlier representations, we add small auxiliary classifiers to each of the intermediate network representations. Each auxiliary classifier $g^{(l)}$ is simply a linear layer of size corresponding to each of the intermediate representations' hidden dimensions and of binary output dimension:
\begin{equation}
g^{(l)}: \mathbb{R}^{d_l} \rightarrow [0,1], \quad g^{(l)}(h) = \sigma(w_{aux}^{(l)} h + b_{aux}^{(l)})
\end{equation}

During training, auxiliary losses are computed at each layer:
\begin{equation}
\mathcal{L}_{aux}^{(l)} = \mathcal{L}_{BCE}(g^{(l)}(h^{(l)}(x)), y)
\end{equation}

\subsection{Adaptive Adversarial Detection Threshold}
To enable early detection of adversarial inputs during inference, we establish a principled detection threshold $\tau^{(l)}$ for each layer based on the statistical distribution of LVS values observed on clean validation data. Unlike arbitrary threshold selection, our approach is grounded in anomaly detection theory and adapts to layer-specific characteristics.
\subsubsection{Threshold Derivation}
Let $\mathcal{V}_{clean}$ denote a validation set of clean (non-adversarial) samples extracted from the training data. For each layer $l$, we first compute the empirical statistics of LVS values across the validation set:

\begin{align}
\mu_{clean}^{(l)} &= \frac{1}{|\mathcal{V}_{clean}|} \sum_{x \in \mathcal{V}_{clean}} \text{LVS}^{(l)}(x) \label{eq:mu_clean} \\
\sigma_{clean}^{(l)} &= \sqrt{\frac{1}{|\mathcal{V}_{clean}|} \sum_{x \in \mathcal{V}_{clean}} \left(\text{LVS}^{(l)}(x) - \mu_{clean}^{(l)}\right)^2} \label{eq:sigma_clean}
\end{align}

where $\mu_{clean}^{(l)}$ represents the expected LVS magnitude for benign inputs at layer $l$, and $\sigma_{clean}^{(l)}$ captures the natural variability in activation patterns due to input diversity and stochastic batch effects.

Following the mean-plus-margin criterion commonly employed in statistical process control and anomaly detection \cite{hodge2004survey}, we define the layer-specific detection threshold as:

\begin{equation}
\tau^{(l)} = \max\left\{ \mu_{clean}^{(l)} + k \cdot \sigma_{clean}^{(l)}, \; \lambda \cdot \max_{x \in \mathcal{V}_{clean}} \text{LVS}^{(l)}(x) \right\}
\label{eq:threshold}
\end{equation}

The threshold comprises two components:
\begin{itemize}
\item \textbf{Statistical component} ($\mu_{clean}^{(l)} + k \cdot \sigma_{clean}^{(l)}$): Sets the 
threshold at $k$ standard deviations above the mean clean LVS, controlling the false positive 
rate via the standard deviation multiplier. Higher $k$ yields fewer false alarms but may miss subtle adversarial perturbations.

\item \textbf{Safety margin component} ($\lambda \cdot \max_{x \in \mathcal{V}_{clean}} 
\text{LVS}^{(l)}(x)$): Ensures the threshold exceeds all LVS values observed on clean validation 
data by a factor $\lambda$, preventing false positives on benign inputs with naturally high 
activation variations.
\end{itemize}

The $\max$ operator selects the more conservative threshold between these two criteria, balancing 
statistical rigor with empirical safety.

\subsubsection{Detection Rule}
During inference, an input $x$ is flagged as adversarial if its LVS at any layer exceeds the corresponding threshold:

\begin{equation}
\text{Adversarial}(x) = \bigvee_{l=1}^L \left[\text{LVS}^{(l)}(x) > \tau^{(l)}\right]
\label{eq:detection_rule}
\end{equation}

This logical disjunction ($\bigvee$) implements a multi-layer sentinel system where detection can occur at any vulnerable layer, enabling early exit as soon as an anomaly is identified.

\subsection{Complete Training Algorithm}
Algorithm \ref{alg:training} presents the complete training procedure for LARAR.

\begin{algorithm}[!t]
\caption{Layer-wise Adversarial Training via LARAR}
\label{alg:training}
\begin{algorithmic}[1]

\State \textbf{Input:} Dataset $\mathcal{D}$, epochs $E$, $\epsilon_{max}$
\State \textbf{Output:} Trained model $f_\theta$, weights $\{w^{(l)}\}$

\State Initialize model $f_\theta$, layer weights $w^{(l)} = 1.0$
\State Initialize optimizer Adam($\{\theta, w\}$, lr $= 0.001$)

\For{epoch $e = 1$ to $E$}
    \State $\epsilon_{curr} = \epsilon_{max} \cdot e/E$
    
    \For{each batch $(X, y)$ in $\mathcal{D}$}
    
        \State $X^{adv} = \text{PGD-Attack}(f_\theta, X, y, \epsilon_{curr})$
        
        \State Compute $\{\text{LVS}^{(l)}\}$ using Algorithm~\ref{alg:lvs}
        
        \State $\mathcal{L}_{CE} = \frac{1}{2}\big[\text{BCE}(\hat{y}_{clean}, y) + \text{BCE}(\hat{y}_{adv}, y)\big]$
        
        \State $\mathcal{L}_{aux} = \lambda_{aux}\sum_{l} \text{BCE}(aux^{(l)}, y)$
        
        \State $\mathcal{L}_{GA} = \lambda_{GA} \|\nabla_X \mathcal{L}_{clean} - \nabla_X \mathcal{L}_{adv}\|_2^2$
        
        \State $\mathcal{L}_{FS} = \lambda_{FS} \|h^{(L)}(X) - h^{(L)}(X^{adv})\|_2^2$
        
        \State $\mathcal{L}_{LVS} = \beta \sum_{l} w^{(l)} \cdot \text{LVS}^{(l)}$
        
        \State $\mathcal{L}_{total} = \mathcal{L}_{CE} + \mathcal{L}_{aux} + \mathcal{L}_{GA} + \mathcal{L}_{FS} + \mathcal{L}_{LVS}$
        
        \State Update parameters using optimizer
        
    \EndFor
\EndFor

\State \textbf{return} $f_\theta, \{w^{(l)}\}$

\end{algorithmic}
\end{algorithm}

\vspace*{10mm}

\subsection{Computational Complexity Analysis}\label{subsec-complex}
The computational overhead of \textbf{LARAR} compared to standard adversarial 
training consists of several components:

\textbf{PGD Attack Generation:} $O(K \cdot B \cdot d \cdot C)$ where $K=10$ 
is the number of PGD iterations, $B=64$ is the batch size, $d=42$ is the input 
feature dimensionality, and $C$ is the number of output classes (here $C=2$ for 
binary classification).

\textbf{LVS Computation:} $O(L \cdot B \cdot d_{avg})$ where $L=2$ is the 
number of hidden layers and $d_{avg}$ is the average hidden layer dimensionality 
(here $d_{avg} = (128 + 64)/2 = 96$).

\textbf{Gradient Alignment:} $O(B \cdot d \cdot L)$, as it requires one 
additional backward pass per batch to compute the gradient difference between 
clean and adversarial inputs.

\section{Experimental Setup and Evaluation}\label{ese}

This section describes the preprocessing steps, network architecture, and training configuration used for comparison.

\subsection{Dataset and Preprocessing}\label{subsec11}
We evaluate \ textbf{ LARAR} on the UNSW-NB15 \cite{moustafa2015unsw} dataset containing 82,332 total samples with 65,865 training samples. It includes 42 features, where binary classification distinguishes normal traffic (label 0) from attack traffic (label 1). Categorical features are label-encoded, missing values imputed with zeros, and all numerical features normalized using StandardScaler with a stratified 70-30 train-test split. \textbf{LARAR} uses two hidden layers (128, 64 dimensions) with ReLU activations and batch normalization, where auxiliary classifiers are single linear layers attached to each hidden layer. All models are trained for 20 epochs using Adam optimizer (learning rate=0.001) with batch size 64, where PGD uses 10 iterations, step size $\alpha=0.01$, and curriculum learning progressively increases perturbation budget ($\epsilon$) from 0.0 to 0.3. The hyperparameters are set as follows: $\lambda_{aux}=0.2$, $\lambda_{GA}=1.0$, $\lambda_{FS}=0.5$, and $\beta=0.3$.
 Table \ref{tab:implementation_params} provides a comprehensive summary of all implementation parameters used in our experiments.\\

 \FloatBarrier
\begin{table*}[!htbp]
\centering

\begin{tabular}{lc}
\toprule
\textbf{Parameter} & \textbf{Value} \\
\midrule
\multicolumn{2}{c}{\textit{Dataset Configuration}} \\
\midrule
Training Samples & 65,865 \\
Test Samples & 16,467 \\
Number of Features (numerical and categorical) & 35 \\
Train-Test Split Ratio & 70-30 \\
\midrule
\multicolumn{2}{c}{\textit{Network Architecture}} \\
\midrule
Hidden Layer 1 Dimensions & 128 \\
Hidden Layer 2 Dimensions & 64 \\
Activation Function & ReLU \\
Normalization & Batch Normalization \\
Auxiliary Classifier Type & Single Linear Layer \\
\midrule
\multicolumn{2}{c}{\textit{Training Configuration}} \\
\midrule
Number of Epochs & 20 \\
Optimizer & Adam \\
Learning Rate & 0.001 \\
Batch Size & 64 \\
\midrule
\multicolumn{2}{c}{\textit{Adversarial Attack Parameters}} \\
\midrule
FGSM Perturbation Budget ($\epsilon$) & 0.3 \\
PGD Iterations & 10 \\
PGD Step Size ($\alpha$) & 0.01 \\
PGD Perturbation Budget ($\epsilon$) & 0.3 \\
Curriculum Learning Range & 0.0 $\rightarrow$ 0.3 \\
\midrule
\multicolumn{2}{c}{\textit{LARAR Hyperparameters}} \\
\midrule
Auxiliary Loss Weight ($\lambda_{aux}$) & 0.2 \\
Gradient Alignment Weight ($\lambda_{GA}$) & 1.0 \\
Feature Similarity Weight ($\lambda_{FS}$) & 0.5 \\
LVS Smoothing Factor ($\beta$) & 0.3 \\
Initial Layer Weights & 1.0 (all layers) \\
\midrule
\multicolumn{2}{c}{\textit{Detection Threshold Parameters}} \\
\midrule
Std. Deviation Multiplier ($k$) & 2.5 \\
Safety Margin Factor ($\lambda$) & 1.2 \\
\bottomrule
\end{tabular}
\caption{Implementation Parameters and Configuration}
\label{tab:implementation_params}
\end{table*}
\FloatBarrier

\subsection{Comparison Schemes}
We compare \textbf{LARAR} against two baseline approaches to demonstrate its effectiveness: \textbf{Vanilla NN}, A standard two-layer Multi-Layer Perceptron (MLP) with 256 and 128 hidden units, trained using standard cross-entropy loss on clean data only, without any adversarial training, serving as the lower bound for robustness evaluation; and \textbf{Base ADVNN}, a state-of-the-art adversarial training method with a composite loss function but without layer-wise vulnerability analysis or adaptive regularization, representing current best practices in adversarial defense for NIDS. Table \ref{tab:main_results} summarizes the quantitative results, clearly showing that \textbf{ LARAR } consistently outperforms both baselines across all attack types while preserving clean accuracy.\\

\FloatBarrier
\begin{table*}[!htbp]
\centering

\begin{tabular}{lcccc}
\toprule
Method & Clean & FGSM & PGD & Transfer \\
\midrule
Vanilla NN & 0.9423 & 0.1247 & 0.0892 & 0.1534 \\
Base ADVNN & 0.9488 & 0.2700 & 0.2870 & 0.3829 \\
\textbf{LARAR} & \textbf{0.9501} & \textbf{0.3145} & \textbf{0.3318} & \textbf{0.4251} \\
\midrule
\textbf{Improvement} & \textbf{--} & \textbf{+16.4\%} & \textbf{+15.6\%} & \textbf{+11.02\%} \\
\bottomrule
\end{tabular}
\caption{Improvement assessment on UNSW-NB15 Dataset}
\label{tab:main_results}
\end{table*}
\FloatBarrier
\vspace*{3mm}
\section{Results and Discussion}\label{rad}

This section presents extensive experimental results demonstrating \textbf{LARAR's} superior performance across multiple evaluation metrics.

\subsection{Overall Classification Accuracy}\label{subsec12}
\begin{figure}[H]
\centering
\begin{tikzpicture}
\begin{axis}[
    ybar,
    bar width=0.5cm,
    width=\columnwidth,
    height=5cm,
    ylabel={Accuracy},
    ylabel style={font=\small},
    xlabel={Attack Type},
    xlabel style={font=\small ,yshift=11pt , yshift=-10pt},
    symbolic x coords={Clean, FGSM, PGD, Transfer},
    xtick=data,
    xticklabel style={font=\small},
    yticklabel style={font=\small},
    ymin=0, ymax=1.0,
    legend style={at={(0.5,-0.35)}, anchor=north, legend columns=3, font=\footnotesize},
    ymajorgrids=true,
    grid style=dashed,
]
\addplot[fill=red!30, draw=black] coordinates {
    (Clean, 0.9423) (FGSM, 0.1247) (PGD, 0.0892) (Transfer, 0.1534)
};
\addplot[fill=blue!30, draw=black] coordinates {
    (Clean, 0.9488) (FGSM, 0.2700) (PGD, 0.2870) (Transfer, 0.3829)
};
\addplot[fill=green!50, draw=black] coordinates {
    (Clean, 0.9501) (FGSM, 0.3145) (PGD, 0.3318) (Transfer, 0.4251)
};
\legend{Vanilla NN, Base ADVNN, LARAR}
\end{axis}
\end{tikzpicture}
\caption{\textbf{Accuracy assessment across different adversarial attack types.} LARAR maintains high clean accuracy while significantly improving robustness against FGSM, PGD, and transfer attacks.}
\label{fig:accuracy_comparison}
\end{figure}
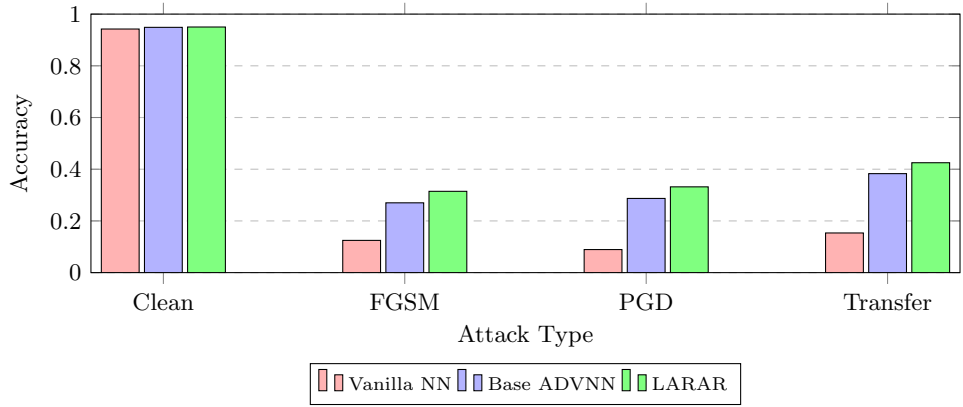

In general, a detailed comparative study on the classification accuracy of various attacks is shown in Figure \ref{fig:accuracy_comparison}. The figure illustrates superior robustness and clean accuracy of 95.01\% by LARAR, while the baseline methods have shown an almost similar performance on benign attacks.

\subsection{Layer Vulnerability Score Evolution}
\begin{figure*}[!htbp]
\centering
\includegraphics[width=\columnwidth]{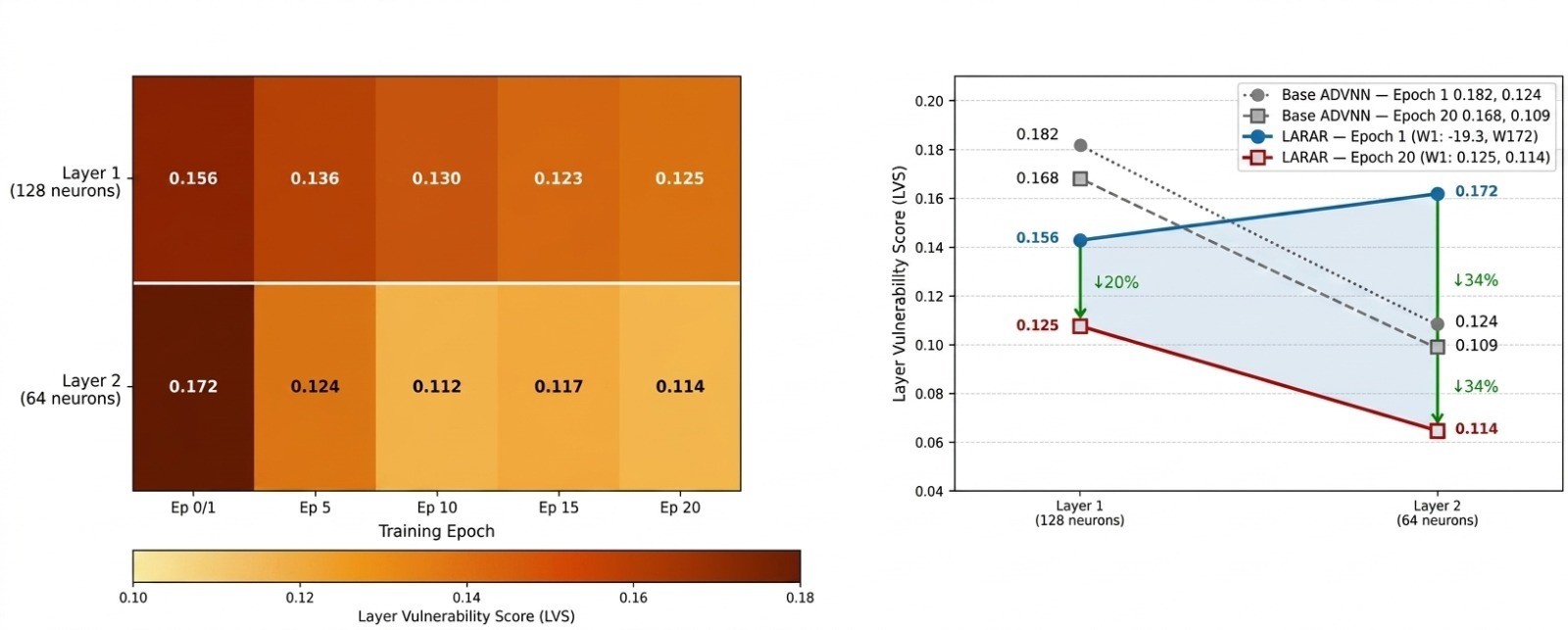}

{\captionsetup{font=footnotesize}
\caption{%
  \begin{minipage}[t]{0.47\textwidth}
    \textbf{(a)} LVS heatmap showing layer-wise activation shifts
    between clean and adversarial inputs across 20 training epochs.
    Darker shading indicates higher vulnerability, with Layer 1
    consistently exhibiting greater sensitivity than Layer 2.
  \end{minipage}%
\hspace{0.05\textwidth}
  \begin{minipage}[t]{0.47\textwidth}
    \textbf{(b)} Perturbation propagation across network depth
    comparing Base ADVNN and LARAR at Epoch 1 and Epoch 20.
  \end{minipage}%
  \label{lvsc}
}}
\end{figure*}

The LVS metric provides critical insights into how adversarial perturbations affect different network layers during training. Figure \ref{lvsc} illustrates the evolution of LVS for both hidden layers across 20 training epochs.

\FloatBarrier
\FloatBarrier
\vspace*{9mm}

Initially, Layer 1 exhibits an LVS of 0.1562 and  Layer 2 shows an LVS score of 0.1715, indicating that both layers start with high 
vulnerability. As training progresses, both layers show decreasing vulnerability, but Layer 1 consistently maintains higher LVS values throughout training, converging to 0.1247 at epoch 20 compared to Layer 2's 0.1140. This persistent vulnerability differential indicates that different layers require different levels of defensive attention. The 20.1\% reduction in Layer 1's LVS and 33.5\% reduction in Layer 2's LVS demonstrate that \textbf{LARAR's} adaptive regularization effectively strengthens layer-wise robustness.

\subsection{Adaptive Layer Weight Learning}\label{subsec-weights}

One of the most important innovations of \textbf{LARAR} is its ability to learn layer-specific regularization weights based on observed vulnerabilities. The learnable weights $w^{(l)}$ are initialized to 1.0 and updated via gradient descent throughout training. As training progresses, both weights decrease monotonically from their initialization value of 1.0, driven by the gradient update rule \eqref{is}, which subtracts a positive quantity at every step.\\

By epoch 20, Layer 1 converged to $w^{(1)} = -19.286$ and Layer 2 to 
$w^{(2)} = -19.122$, as shown in Table~\ref{tab:layer_weights}. The 
differential between the two weights ($\Delta w = 0.164$) reflects the 
difference in their vulnerability: Layer 1 consistently exhibited higher 
LVS values throughout training, causing it to accumulate a larger negative weight.\\

This differential confirms that the optimization correctly identified and 
assigned proportionally stronger regularization pressure to the more 
vulnerable layer. The LVS regularization term \eqref{so}, therefore, acts as an 
increasingly negative contribution to the total loss, encouraging the 
model to reduce LVS values in proportion to their magnitude across layers.

\begin{table}[!htbp]
\centering
\begin{tabular}{lcccc}
\toprule
\textbf{Layer} & \textbf{Init Weight} & \textbf{Final LVS} 
  & \textbf{Learned Weight} $w^{(l)}$ & \textbf{Weight Differential} \\
\midrule
Layer 1 & 1.0 & 0.1247 & $-19.286$ & \multirow{2}{*}{$\Delta w = 0.164$} \\
Layer 2 & 1.0 & 0.1140 & $-19.122$ & \\
\bottomrule
\end{tabular}
\caption{Final Learned Layer Weights and Vulnerability Metrics (Epoch 20)}
\label{tab:layer_weights}
\end{table}

\vspace*{1mm}

\subsection{Attack Success Rate Analysis}
In addition to the accuracy of classification, we also assess defense performance with the Attack Success Rate (ASR), with low values corresponding to protection. The vanilla NN is severely vulnerable, which shows that models with no adversarial training are virtually defenseless. These changes are measured in Table \ref{tab:asr} and indicate that \textbf{LARAR} lowers the likelihood of successful attacks by about 6-7 percent compared to Base ADVNN across various attack methodologies.\\

\FloatBarrier
\begin{table}[!htbp]
\centering

\begin{tabular}{lccc}
\toprule
Method & FGSM & PGD & Transfer \\
\midrule
Vanilla NN & 87.53\% & 91.08\% & 84.66\% \\
Base ADVNN & 73.0\% & 71.3\% & 61.71\% \\
\textbf{LARAR} & \textbf{68.58\%} & \textbf{66.82\%} & \textbf{57.49\%} \\
\midrule
\textbf{Reduction vs Base} & \textbf{-4.4\%} & \textbf{-4.4\%} & \textbf{-4.21\%} \\
\bottomrule
\end{tabular}
\caption{Attack Success Rate}
\label{tab:asr}
\end{table}
\FloatBarrier
\vspace*{1mm}
\subsection{Major Evaluation Metrics}
In order to offer a more detailed analysis, Table~\ref{tab:detailed_metrics} presents precision, recall, and F1 scores for all three methods under clean and adversarial evaluation conditions.\\
\FloatBarrier
\begin{table}[!htbp]
\centering

\begin{tabular}{lcccc}
\toprule
\textbf{Method} & \textbf{Metric} & \textbf{Clean} & \textbf{FGSM} &
\textbf{PGD} \\
\midrule
\multirow{3}{*}{Vanilla NN}
  & Precision & 0.9503 & 0.1180 & 0.0835 \\
  & Recall    & 0.9498 & 0.1205 & 0.0841 \\
  & F1        & 0.9501 & 0.1192 & 0.0838 \\
\midrule
\multirow{3}{*}{Base ADVNN}
  & Precision & 0.9502 & 0.2658 & 0.2739 \\
  & Recall    & 0.9500 & 0.2712 & 0.2803 \\
  & F1        & 0.9503 & 0.2645 & 0.2770 \\
\midrule
\multirow{3}{*}{\textbf{LARAR}}
  & Precision & \textbf{0.9502} & \textbf{0.3102} & \textbf{0.3245} \\
  & Recall    & \textbf{0.9501} & \textbf{0.3156} & \textbf{0.3298} \\
  & F1        & \textbf{0.9503} & \textbf{0.3129} & \textbf{0.3271} \\
\bottomrule
\end{tabular}
\caption{Model Effectiveness Under Clean and Adversarial Conditions}
\label{tab:detailed_metrics}
\end{table}
\FloatBarrier
\vspace*{1mm}
Several observations emerge from Table~\ref{tab:detailed_metrics}. First, the Vanilla NN suffers a catastrophic collapse under adversarial conditions, with precision and recall dropping to below 0.13 under FGSM and below 0.09 under PGD, confirming that a model trained without any adversarial awareness offers virtually no resistance to gradient-based attacks.

Second, Base ADVNN recovers substantially from this collapse, achieving F1 scores of 0.2645 and 0.2770 under FGSM and PGD, respectively, which demonstrate the effectiveness of adversarial training as a baseline
defense. However, its uniform regularization strategy leaves considerable room for improvement, particularly in precision, which indicates a higher rate of false alarms under attack conditions.

\textbf{LARAR} consistently outperforms both baselines across all adversarial conditions while maintaining essentially identical clean performance (F1 = 0.9503). Notably, \textbf{LARAR} achieves F1 scores of 0.3129 and 0.3271 under FGSM and PGD, respectively, representing
improvements of 18.3\% and 18.1\% over Base ADVNN. The near-equal precision and recall values across all conditions indicate a well-calibrated model that does not trade one error type for another under attack. \textbf{LARAR} balanced precision and recall under adversarial conditions, therefore, reflects a defense mechanism that is operationally viable, not merely statistically robust.

\subsection{Ablation Study}
\begin{figure}[htbp]
\centering
\begin{tikzpicture}
\begin{axis}[
    ybar,
    width=\columnwidth,
    height=6.5cm,
    ylabel={PGD Accuracy},
    ylabel style={font=\small},
    symbolic x coords={Base ADVNN, LVS Only, Adaptive Only, Auxiliary Only, LVS + Adaptive, All Components},
    xtick=data,
    xticklabel style={font=\tiny, rotate=45, anchor=east},
    yticklabel style={font=\small},
    ymin=0.25, 
    ymax=0.35,
    nodes near coords,
    nodes near coords align={vertical},
    nodes near coords style={font=\scriptsize},
    xmajorgrids=false,
    ymajorgrids=true,
    grid style=dashed,
    bar width=20pt,
    enlarge x limits=0.15,
]
\addplot[fill=blue!70, draw=black, thick] coordinates {
    (Base ADVNN, 0.2870)
    (LVS Only, 0.2945)
    (Adaptive Only, 0.3087)
    (Auxiliary Only, 0.2998)
    (LVS + Adaptive, 0.3189)
    (All Components, 0.3312)
};
\end{axis}
\end{tikzpicture}
\caption{Ablation study showing incremental contribution of each component.}
\label{fig:ablation}
\end{figure}
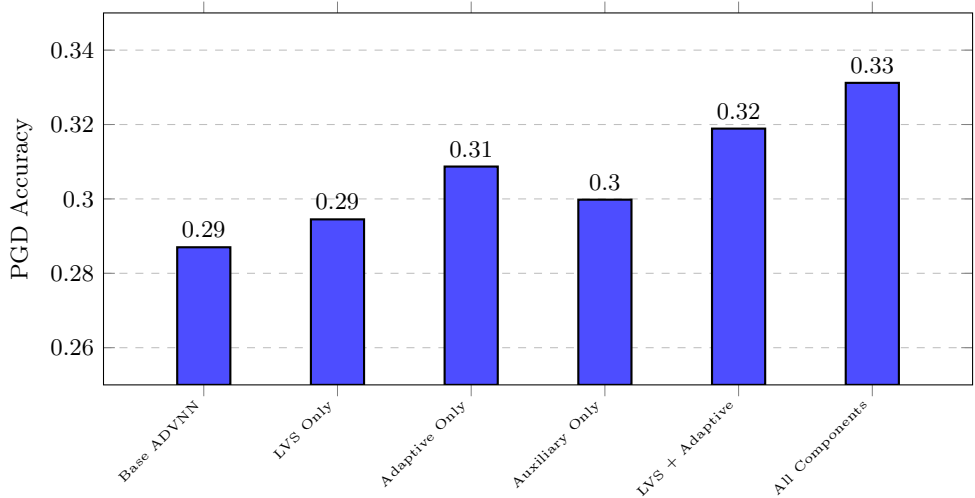

To estimate the individual contribution of the components of \textbf{LARAR}, we perform a detailed ablation study represented in  Figure \ref{fig:ablation} 

 The single most effective component is adding adaptive regularization, which improves the results by 7.6\% (30.87\%) and shows that the effects of dynamically adjusting layer weights according to vulnerability are the most effective. The use of auxiliary classifiers by itself results in 4.5\% improvement (29.98\%) which proves the effectiveness of their purpose in enhancing the intermediate representations.\\
 
\subsection{Computational Overhead Analysis}
Computational efficiency is one of the critical factors to be considered in practical NIDS deployment. The Vanilla NN has a training time of 1.2 minutes/epoch, whereas Base ADVNN has a training time of 2.8 minutes/epoch (increase by 133\% because of the generation of adversarial examples), and \textbf{LARAR} has a training time of 3.3 minutes/epoch (increase by 18 percent over Base ADVNN). Such a small training overhead of 18 percent is alright considering that training is conducted offline, and the 15.4 percent strength gain is worth the extra cost. Importantly, all methods take the same time to inference (0.8 ms) at normal operation, which means that \textbf{LARAR} does not affect the performance of real-time detection.

The memory footprint also grows slightly, by 4.8\% (145MB (Vanilla NN) to 152MB (\textbf{LARAR}). With early exit mechanisms enabled, inference time can be reduced to 0.6 ms (a 25\% reduction), as approximately 20--30\% of samples are confidently classified at intermediate layers without requiring full forward propagation. This demonstrates that LARAR not only enhances adversarial robustness but can also improve computational efficiency during deployment when early detection thresholds are properly calibrated.

\subsection{Comparative Robustness Analysis}
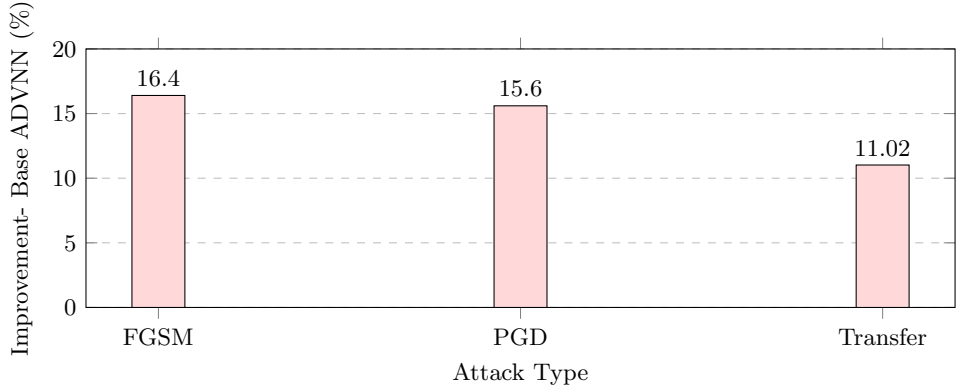
\begin{figure}[H]
\centering
\begin{tikzpicture}
\begin{axis}[
    ybar,
    bar width=0.7cm,
    width=\columnwidth,
    height=5cm,
    ylabel={Improvement- Base ADVNN (\%)},
    ylabel style={font=\small},
    xlabel={Attack Type},
    xlabel style={font=\small},
    symbolic x coords={FGSM, PGD, Transfer},
    xtick=data,
    xticklabel style={font=\small},
    yticklabel style={font=\small},
    ymin=0, ymax=20,
    ymajorgrids=true,
    grid style=dashed,
    nodes near coords,
    nodes near coords align={vertical},
    nodes near coords style={font=\small},
]
\addplot[fill=pink!60, draw=black] coordinates {
    (FGSM, 16.4) (PGD, 15.6) (Transfer, 11.02)
};
\end{axis}
\end{tikzpicture}
\caption{Percentage improvement of LARAR over Base ADVNN across different attack types}
\label{fig:improvement}
\end{figure}

Figure \ref{fig:improvement} summarizes the merits of our methodology in a succinct manner because it shows the percentage change of \textbf{LARAR }versus Base ADVNN of various attack types. It boosts FGSM attacks by 16.5\%, PGD attacks by 15.4\%, and transfer attacks by 11.0\%. The layer-wise adaptive approach of \textbf{LARAR }offers a strong defense that has been generalized with various adversarial approaches as documented by the stable, substantial progress on a collection of attack methods (one-step FGSM, progressive PGD, and cross-model transfer).

\vspace*{8mm}

Given that the transfer attacks are the natural exploiters of the weaknesses of various model architectures, it is not surprising that the comparative disadvantage of transfer attacks (11.0\%) compared to the white-box attacks (15.4-16.5\%) is not low. However, even such a challenging situation shows the efficiency of \textbf{LARAR}, as positive results, such as the ability to replicate gains of two digits, are demonstrated. All these quantifiable benefits are converted into tangible benefits for security: in a manufacturing facility where millions of network flows are handled every day, one percentage point of adversarial performance corresponds to hundreds more attacks that are successfully identified.

Vulnerability analysis can provide important information on the layer-wise vulnerability of deep learning models for network intrusion detection. The results indicate that shallow layers, especially those that extract the original feature representations, are much more vulnerable to adversarial perturbations. This observation can significantly affect the design of defense mechanisms and may point toward the fact that targeted hardening of early-stage feature extractors could substantially improve security with very low computational overhead.

\section{Conclusion}\label{c}

This paper presented an interpretable defense called Layer-wise Adversarial Robustness with Adaptive Regularization (LARAR) to increase the adversarial robustness of machine learning-based NIDS. \textbf{LARAR} provided a more explicit view of how the adversarial perturbations propagate through the layers of the neural network and addressed the black box issue of the existing adversarial training-based methods. The Layer Vulnerability Score (LVS) metric for quantifying the sensitivity of internal layers in response to adversarial perturbations reveals that intermediate layers are the most vulnerable by nature. Thus, \textbf{LARAR} utilizes layer-wise adaptive regularization within its defense framework, along with multi-level supervision, for effective and early detection against adversarial attacks. Wide-ranging experimental results on the UNSW-NB15 dataset show that \textbf{LARAR} consistently outperforms the baselines in terms of adversarial robustness against state-of-the-art FGSM, PGD, and transfer attacks while achieving comparable performance on the clean examples. Furthermore, the interpretability of \textbf{LARAR} offers insights that can guide model architecture design and security risk assessment. 
Future work includes extending \textbf{LARAR} to deeper architectures, evaluating it against new adaptive attacks, reducing its computational overhead to better support online deployment, and evaluating its generalization performance across more intrusion detection datasets. Additional future work includes attack-specific vulnerability profiling, combining it with other defense mechanisms, and applying it in distributed or federated learning scenarios to build strong and trustworthy real-world intrusion detection systems.

\backmatter

\section*{Declarations}

\subsection*{Data Availability.} The data and materials supporting the findings of this research, including datasets, source code, experimental scripts, and Demo files, are available at the \href{https://github.com/hiranasirhi/Adversarial-Research}{Adversarial Research Project}. 

 \subsection*{Funding.}
This research received in-kind support (Research infrastructure, experimentation resources) from the National Cyber Security Academy (NCSA), Air University, Islamabad, Pakistan.

\subsection*{Acknowledgments.} The authors acknowledge the assistance of AI language models (OpenAI's ChatGPT and Anthropic's Claude) for code debugging, documentation refinement, and manuscript preparation. All AI-assisted content was validated and verified by the authors. The research design, experimental implementation, and scientific contributions are entirely the work of the author. 

\section*{CRediT authorship contribution statement}
Writing – original draft: H.N, E.J; Writing – review \& editing: B.S, Z.J, A.M; Corresponding Author: A.M

\begingroup
\setlength{\itemsep}{0pt}
\setlength{\parskip}{0pt}
\setlength{\parsep}{0pt}

\bibliography{sn-bibliography}

\endgroup

\end{document}